\documentclass [prl,twocolumn,showpacs,preprintnumbers,amsmath,amssymb]{revtex4}

\usepackage{graphicx}
\usepackage{dcolumn}

\newcommand{\beq}{\begin{equation}}
\newcommand{\eeq}{\end{equation}}
\newcommand{\beqa}{\begin{eqnarray}}
\newcommand{\eeqa}{\end{eqnarray}}

\begin{document}
\title{Electron-phonon Interaction close to a Mott transition}

\author{G. Sangiovanni$^1$, M. Capone$^{1,2}$, C. Castellani$^1$,
 and M. Grilli$^1$} 

\affiliation{$^1$INFM Statistical Mechanics and Complexity Center, 
and Dipartimento di Fisica\\
Universit\`a di Roma "La Sapienza" piazzale Aldo Moro 5, I-00185 Roma, Italy}
\affiliation{$^2$Enrico Fermi Center, Roma, Italy}

\begin{abstract}
The effect of Holstein electron-phonon interaction on a Hubbard model 
close to a Mott-Hubbard transition at half-filling is investigated by means of
Dynamical Mean-Field Theory. We observe a reduction of the effective mass 
that we interpret in terms of a reduced effective repulsion. 
When the repulsion is rescaled to take into account this effect,
the quasiparticle low-energy features are unaffected by the 
electron-phonon interaction.
Phonon features are only observed within the high-energy Hubbard bands.
The lack of electron-phonon fingerprints in the quasiparticle physics
can be explained interpreting the quasiparticle motion in terms of rare 
fast processes.
\end{abstract}
\date{\today}
\maketitle

Electron-phonon (e-ph) interaction is a key effect in solids, that gives
rise to superconductivity, charge-density-waves and other important 
effects. While the physical phenomena caused by e-ph coupling
in weakly correlated materials are reasonably well understood, much less
is known about the role of e-ph coupling in strongly
correlated materials, i.e., in compounds in which the short-range
Coulomb repulsion is larger than the typical kinetic energy.
This issue, besides its obvious general interest, has been strongly
revived by recent experiments which suggest a role of lattice effects 
in the high-temperature superconducting cuprates (which are
known to be strongly correlated compounds) \cite{lanzaraealtri}, 
and by the realization that strong correlations are important in 
phonon-mediated superconductors as the fullerenes \cite{fullerene,capone}.

The interplay between  e-ph
and electron-electron (e-e) interactions is quite a complicated problem, 
in which a variety of physical regimes and different phases can be
realized.
In the present paper we focus on the strongly correlated regime, in which
the e-e interaction is assumed to be quite large, and the mobility of the
electrons is strongly reduced. Our main concern is to understand whether 
and to which extent the e-ph coupling can still influence the electronic
properties in this regime.
We notice that the symmetry of the e-ph coupling is important
in this regard: It has indeed been shown that a Jahn-Teller coupling, 
where the phonon variables are coupled  with orbital and spin degrees of freedom is weakly
affected by Hubbard repulsion $U$, whereas a Holstein coupling with the local
charge is strongly suppressed \cite{gunnarsson,capone}.
Nevertheless, even the Holstein coupling, if sizeable,  
can have a qualitative physical effect in a correlated material. 
For instance, it can induce
phase separation close to the density-driven Mott transition \cite{caponeesoci}.
A recent important contribution comes from the two-dimensional 
Quantum Monte Carlo study of Ref. \cite{scalapino},
where a small $U$ is found to suppresses the e-ph coupling at 
all electron and phonon momenta, while, in the strongly correlated regime, the 
forward-scattering amplitude due to e-ph processes substantially 
increases with $U$, leading to a non-monotonic behavior of the 
effective coupling.

In this work we consider again the simplest case of the Holstein
interaction, whose formal simplicity allows for extensive systematic studies.
Our Hamiltonian reads
\begin{eqnarray}
H = & -t\sum_{\langle i,j\rangle,\sigma}  c^{\dagger}_{i,\sigma} c_{j,\sigma} 
+ U \sum_i n_{i\uparrow} n_{i\downarrow} \nonumber\\
& - g\sum_i n_i(a_i +a^{\dagger}_i) + \omega_0 \sum_i a^{\dagger}_i a_i,
\label{hamiltonian}
\end{eqnarray}
where $c_{i,\sigma}$ ($c^{\dagger}_{i,\sigma}$) and $a_i$ ($a^{\dagger}_i$) are, respectively,
 destruction (creation) operators for 
fermions with spin $\sigma$ and for local vibrations of frequency $\omega_0$ 
on site $i$, $t$ is 
the hopping amplitude, $U$ is the local Hubbard repulsion and $g$ is an e-ph coupling.
Despite its simplified character, the Hubbard-Holstein model is expected
to display a complicated phase diagram as a function of its parameters,
the correlation strength $U/D$ ($D$ is the electronic half-bandwidth), 
the e-ph coupling 
$\lambda=2g^2/\omega_0D$, the adiabatic ratio $\omega_0/D$ and the density 
$n$ \cite{hewson,jeon,caponeesoci}.
As notable examples, this model presents the physics of the Mott transition
when the Hubbard repulsion is the most important scale \cite{revdmft}, 
a crossover to polaronic carriers or a bipolaronic metal-insulator transition 
for small $U$ and large $\lambda$ \cite{caponeciuchi,hewson}, and it is expected to be unstable toward superconductivity, charge-density waves and antiferromagnetism in 
different regions of the phase diagrams \cite{ff}.
To gain physical insight into some aspects of this phase diagram, it is 
necessary to fix some parameters, and put ourselves in a given regime.
As we have said, we investigate the strongly correlated metallic region,
close to the Mott transition at half-filling. Our analysis is therefore 
relevant for metallic phases of correlated materials, such as the doped 
cuprates.

We solve the model by means of Dynamical Mean-Field Theory (DMFT), which
has emerged as one of the most reliable tools from the analysis of both
correlated materials, and e-ph interactions. The method maps the lattice
model onto a local theory which still retains full quantum dynamics, and
it is therefore expected to be quite accurate for models with local 
interactions, such as (\ref{hamiltonian}). The DMFT solution is enforced
by solving the local dynamical theory and imposing a self-consistency
condition \cite{revdmft}. The local problem is equivalent to an 
Anderson-Holstein impurity model\cite{hewsonam} 
which has to be solved with some tool. 
In our work we use exact
diagonalization \cite{caffarel}, truncating the infinite phonon Hilbert
space allowing up to $N_{max}$ phonon states (ranging from $20$ to $40$), 
and using $N_b =9$ sites in the conduction bath.

DMFT has been extensively used to study the half-filled Hubbard model,
showing that, in the  metallic region close to the Mott
transition, the spectral function presents three features: the 
high-energy Hubbard bands at energy $\pm U/2$ and
a narrow resonance at the Fermi level, whose width is proportional to 
the quasiparticle residue $Z=(1-\partial\Sigma(\omega)/\partial\omega\vert_{\omega=0})^{-1}$,
a quantity which decreases for increasing $U$ and vanishes at the Mott
transition. Due to the momentum independence of the self-energy,
$Z$ is also the inverse of the effective mass $Z = m/m^*$.
On general grounds, we can expect two main effects of e-ph 
coupling starting from the situation described above. 
The first is a reduction of $Z$ (equivalent to an 
enhancement of the effective mass). This effect is responsible for the 
polaron crossover for weak repulsion. 
The second effect is a phonon-mediated retarded attraction
between electrons with opposite spins, which opposes the Hubbard repulsion
leading to an effective dynamical interaction
$U_{eff}(\omega)= U - \frac{2g^2\omega_0}{\omega_0^2 -\omega^2}$.
In the antiadiabatic limit $\omega_0 \to \infty$ the attraction becomes
static and equals the bipolaronic binding energy $2g^2/\omega_0$.
While the first effect opposes to the electronic motion, the second 
effect instead favors the motion by screening the Hubbard repulsion 
and leads to an enhancement of $Z$
(at least as far as $U > 2g^2/\omega_0$). The overall effect of the
e-ph coupling on the correlated metallic phase 
is therefore hard to predict on intuitive grounds.
\begin{figure}[htbp]
\begin{center}
\includegraphics[width=7.5cm]{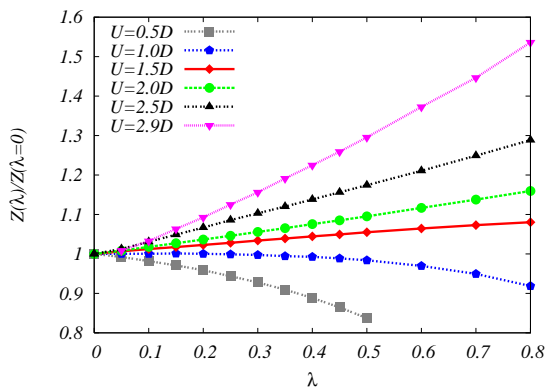}
\end{center}
\caption{(Color online) Effect of e-ph interaction on the quasiparticle weight
$Z$. The ratio 
$Z(\lambda,U)/Z(0,U)$ is shown for different values of $U$ and $\omega_0=0.2D$.
}
\label{figzeta}
\end{figure}

The DMFT study of model (\ref{hamiltonian}) helps us in understanding 
which of the two effects prevails. In Fig. \ref{figzeta} we report the 
quasiparticle
weight $Z$ as a function of $\lambda$ for different values
of $U/D$ for the half-filled Hubbard-Holstein model in the 
relatively adiabatic regime $\omega_0/D = 0.2$. 
In the absence of Hubbard repulsion, the effect of the e-ph
coupling is quite naturally to increase the electron effective mass, and 
decrease $Z$, eventually reaching the bipolaronic metal-insulator
transition \cite{caponeciuchi}. Then we turn on the Hubbard repulsion, 
and plot the ratio $Z(U,\lambda)/Z(U,0)$, in order to disentangle the
effect of the e-ph interaction. 
While $Z(U,\lambda)$ is a decreasing function of $U$ for each $\lambda$,
the above ratio displays a richer behavior. Increasing the value of $U$, 
the effect of the e-ph interaction becomes weaker, until 
a value of $U\simeq 1.5D $ is reached, for which
the e-ph coupling has the surprising effect to increase $Z$
(reduce the effective mass), in agreement with the results found with a
Numerical Renormalization Group solution of DMFT in Ref.\onlinecite{hewson}. 
This naively surprising result can be attributed, in light of the previous
discussion, to the reduction of the effective repulsion, which 
prevails over the polaronic renormalization of the hopping.
In the following we empirically determine an effective {\it static} repulsion 
$U_{eff}(U,\lambda,\omega_0)$,
as the value of $U$ for which a pure Hubbard model has the same
$Z$ as the full Hubbard-Holstein model. 
Since in the antiadiabatic limit
$U_{eff}=U-\lambda D$, we try to parameterize the effective repulsion
as $U_{eff}=U-\eta\lambda D$, where $\eta$ is a dimensionless free parameter 
that we determine by fitting directly $Z(U,\lambda)$ with this functional form.
 
\begin{figure}[htbp]
\begin{center}
\includegraphics[width=7.5cm]{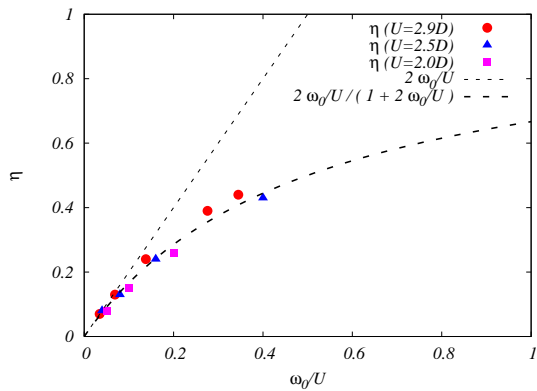}
\end{center}
\caption{(Color online) The coefficient $\eta$ which measures the effective reduction
of the Hubbard repulsion according to $U_{eff}=U-\eta\lambda D$, plotted
as a function of $\omega_0/U$. Different symbols refer to different values of 
$U$. The values of $\lambda$ used for the fitting procedure described 
in the text, range typically from $0$ to $1$.}
\label{figeta}
\end{figure}
Computing $\eta$ for different values of $U$ and $\omega_0$, we find 
the behavior shown in Fig. \ref{figeta}. 
For relatively small $\omega_0/U$, we find that $\eta$ has a linear behavior
in $\omega_0/U$, and bends down for larger phonon frequency closely 
following the functional form $\eta=2\omega_0/U/(1+2\omega_0/U)$
and eventually reaching the asymptotic value $\eta =1$.
We will see in the following that this functional form can be 
derived from an expansion of the effective Kondo coupling\cite{grempel}.

\begin{figure}[htbp]
\begin{center}
\includegraphics[width=8.0cm]{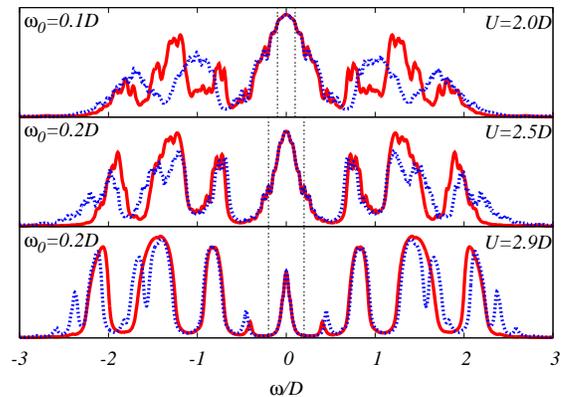}
\end{center}
\caption{(Color online) The electron spectral function for the 
Hubbard-Holstein model 
(dashed line), compared with a pure Hubbard model with  
$U=U_{eff}=U-\eta\lambda D$ (solid line), for different values of 
$U$ and $\omega_0$ (the dot-dashed vertical line indicates $\omega_0$). 
In the two upper panels we take $\lambda=1$, in the lowest $\lambda=0.5$.}
\label{figspettro}
\end{figure}

The analysis of the quasiparticle weight suggests that the effect of 
e-ph interaction on the strongly correlated metallic phase of the Hubbard
model is a partial screening of the repulsion, and the degree of screening
is controlled by $\omega_0/U$ when this quantity is not too large. 
This latter factor contains the information about the lattice dynamics
in an effective way. Of course, at this level, this result could sound
of little physical significance, since it basically arises by a simple
fit procedure. Nonetheless, $Z$ also measures the width and weight
of the low-energy feature in the single-particle spectral function. 
Thus, it is tempting to compare the spectral function for the Holstein 
model with the correspondent quantity for the equivalent Hubbard model
with reduced repulsion. 
As far as $U \gg \omega_0$, we find that the low-energy part of the 
spectrum is basically identical in the two models, while differences 
develop in the high-energy part, where phonon satellites at energies
of the order $\omega_0$ appear in the system coupled to phonons. This 
can be seen in Fig.\ref{figspettro}, where the spectra for three different 
cases (see caption) are shown. 
Moreover, even if the effect of phonons is visible in the high-energy
Hubbard bands, their position is more or less coincident in the 
Hubbard-Holstein model and in the effective Hubbard model. 
This is quite surprising, since the rescaling of $U$ has been derived only
be requiring the same value for the quasiparticle weight.

This findings strongly suggest that our reduction of the dynamical
effective interaction to a static quantity is not a mere fitting, but
it unveils some basic physics.
We can gain some understanding of this scenario by exploiting the 
fact that the quasiparticle peak is associated within DMFT to a Kondo
effect in the impurity model.
It has been recently pointed out that, at the leading order in $1/U$,
the Kondo coupling $J_K$ is only weakly affected by e-ph
coupling \cite{grempel}. This result is closely related to 
a previous calculation addressing the effect of e-ph interaction
on the superexchange coupling in a Hubbard model \cite{nostroj}.
In both cases, the coupling (Kondo or antiferromagnetic) arises in fact, 
at the leading order in $1/U$, from virtual processes in which doubly 
occupied sites are created, with and energy cost of order $U$.
In the presence of an e-ph coupling, the intermediate states
can have an arbitrary number of phonons, but, in the limit in which 
$\omega_0 \ll U$, the corrections are small, and $J_K$ is not
substantially affected. 
More specifically, summing the contributions from all the 
states with different phonon occupation 
(see, e.g., Eq.(8) of Ref.\onlinecite{grempel}), one obtains a 
small correction of $J_K$ given, at first order in  $\lambda_0/U$, by 
\begin{equation}
\label{jkondo}
\frac{J_K(\lambda)}{J_K(0)} \simeq 
1 + \frac{\lambda D}{U}\frac{\frac{2\omega_0}{U}}{1+\frac{2\omega_0}{U}},
\end{equation}
which at first order in $\omega_0/U$ becomes $J_K(\lambda)/J_k(0) \simeq
1+2\omega_0\lambda D/U^2 = 1+4g^2/U^2$.
Eq. (\ref{jkondo}) determines an effective static $U$ for the Kondo problem given by
$U_{eff}^K=U- \eta^K\lambda D$, with $\eta^K= (2\omega_0/U)/(1+2\omega_0/U)$.
The agreement between the $\eta$ extracted from DMFT and this prediction
is evident from Fig. \ref{figeta}.
This provides a strong physical interpretation of our previous 
finding about the effective $U$ giving rise to 
the renormalization of the quasiparticle weight for small $\omega_0/U$,
for which we obtain the same dependence (see Fig. \ref{figeta}).
Indeed within DMFT, the formation of the quasiparticle peak 
can be associated to a Kondo effect, which may in turn
be described as arising from virtual processes of the same kind as the
ones that lead to the Kondo coupling. But for the overall rescaling of 
$U_{eff}$, looking in detail to the phonon-induced modifications
of the low-energy spectrum (the Kondo resonance in the single-impurity
language) we found that these modifications are quantitatively extremely small
and almost invisible on the resolution scale of Fig.\ref{figspettro} \cite{nota2site}.

The picture that emerges from the previous arguments can be summarized
as follows: Quasiparticle motion arises from virtual processes in which 
doubly occupied sites are created. Obviously, these processes are not
so frequent, since the energy scale involved is large, but they are 
extremely rapid (the associated time scale is $\propto 1/U$), 
and consequently are poorly affected by phonon
excitations with a characteristic time scale $1/\omega_0 \gg 1/U$. 
When the phonon frequency is small with respect to $U$, 
the phonon degrees are frozen during the virtual excitation processes.
Therefore, despite the overall electron motion is quite slow
due to the small number of virtual processes 
(which is reflected by the large effective mass), the e-ph
interaction has no major effect except for a slight reduction of the
total static repulsion.

Finally, we come to some recent experimental evidences of phononic
effects in the strongly correlated cuprate superconductors. 
It must be noted that our results refer to the half-filled
case for $U$ smaller than the Mott transition point, while
the experiments obviously refer to a doped system most likely with a 
larger $U$. 
Nevertheless it is tempting to assume that also in the $U$ {\it vs.} 
doping region, relevant for the cuprates, the main low-energy effect of phonons is a small
down-shift of the repulsion, that will be of no relevance when $U$ is
appreciably larger than its value at the Mott transition point.
Our main finding of a completely different effect of e-ph 
coupling on the low-energy part of the spectrum with respect to the high-energy region, 
may provide an explanation of the observation of e-ph interaction 
in the high-energy branch of the ARPES spectra, in contrast to the weak
effect on the low-energy part, as observed in recent isotope effect 
measurements \cite{lanzaranew}.
On the other hand, if the system is close to the Mott transition, a
slight modification of $U_{eff}=U-2\omega_0\lambda D/U$ through the
variation of $\omega_0$ can significantly change the effective mass,
leading to important isotope effects, affecting the penetration
depth, as found in various underdoped samples \cite{keller}.
A more detailed analysis of the evolution of the isotope effects
in the full $U$ {\it vs.} doping diagram is worth being carried out
to provide a more reliable comparison with the available experiments.

A main limitation of our work, in light of previous studies of the problem of
e-ph interactions, is the inability of DMFT to capture the
momentum dependence of the phonon vertex. 
Even for the Holstein model, in which the bare coupling is local, i.e., momentum
independent, it has indeed been shown that correlation effects tend to favor
small exchanged momenta (forward) scattering \cite{scalapino,vertici}.
In this perspective, our DMFT results indicating that the local effects of the
Holstein coupling are strongly renormalized down by correlations,
allow to conclude that ``standard'' features of the e-ph
coupling (polaron crossover, ordinary BCS pairing) are negatively influenced
by correlation, while non-standard effects related to the strong forward
scattering can survive and even get enhanced.
In other words, if any Holstein-like e-ph interaction turns out to be 
relevant in strongly correlated materials,
it must display anomalous features associated to the relevance
of forward scattering and consequently requires a wider framework than 
ordinary Migdal-Eliashberg theory.
On the other hand, the strong reduction by $U$ does not apply to
e-ph coupling involving degrees of freedom non competing with
local charge fluctuations. This is the case for the Jahn-Teller coupling
considered in Ref.\cite{capone}, where the phonon-mediated
pairing is indeed enhanced by strong correlations.

We have studied the half-filled Hubbard-Holstein model in the strongly
correlated regime, where correlation effects are most important and determine 
a clear separation between low-energy quasiparticle features and ``insulating''
Hubbard bands. In this regime, we find that e-ph interaction has basically no
effect on the quasiparticle features as long as the Hubbard repulsion is
suitably rescaled. An effective static repulsion $U_{eff} = U - 2\omega_0\lambda D/U$
in fact reproduces the low-energy features of the Hubbard-Holstein model
using a simple Hubbard model. 
On the other hand, the high-energy Hubbard bands display phononic features
that can not be found in an effective purely electronic model, even if the
same scaling of $U$ also reproduces the correct position of the bands.
The ``protection'' of the low-energy physics with respect to e-ph
interactions is provided by the (small) effect of lattice coupling
on the Kondo coupling and on the resonance shape. We understand this result
in terms of freezing of phonons on the time-scale relevant to virtual
processes underlying the quasiparticle motion.

We are glad to thank E. Tosatti for an enlightening discussion about the
physical content of our results. 
We also acknowledge fruitful discussions with 
E. Cappelluti, S. Ciuchi, M. Fabrizio, and A. Toschi, 
as well as financial support from MIUR Cofin 2003.


\begin{thebibliography}{99}
\bibitem{nota2site} From an analytic solution of the two-site case ($N_b=1$)
we find that these modifications scale with a coupling
$(\omega_0/U)\lambda \ll \lambda$.
\end{thebibliography}
\end{document}